\documentclass[twocolumn]{ceurart}

\usepackage[T1]{fontenc}
\usepackage[utf8]{inputenc}
\usepackage{lmodern}
\usepackage{microtype}
\usepackage{amsmath}
\usepackage{amssymb}
\usepackage{amsfonts}
\usepackage{textcomp}
\usepackage{booktabs}
\usepackage{longtable}
\usepackage{array}
\usepackage{tabularx}
\usepackage{adjustbox}
\usepackage{makecell}
\usepackage{graphicx}

\begin{document}

\copyrightyear{2026}
\copyrightclause{Copyright for this paper by its authors.
  Use permitted under Creative Commons License Attribution 4.0
  International (CC BY 4.0).}

\conference{RecSys in HR '26: The 6th Workshop on Recommender Systems for Human Resources, in conjunction with the 20th ACM Conference on Recommender Systems, September 28--October 2, 2026, Minneapolis, MN, United States.}

\title{Fine-Tuning Strategy, Not Just Data: A Statistically Rigorous Audit of Gender and Stereotypical Bias in Embedding Models for HR Recommendation Systems}

\author[1]{Ekta Yadav}[%
  email=ekta.yadav@randstadusa.com,
]
\fnmark[1]

\address[1]{Randstad N.V.}

\fntext[1]{The views, opinions, and findings expressed in this paper are solely those of the author and do not necessarily reflect the official views, positions, or products of Randstad N.V. No internal Randstad data or production models were evaluated in this research.}

\begin{abstract}
Embedding models are the backbone of modern job- and candidate-matching recommender systems: a resume, a job description, or a query is encoded into a dense vector and recommendations are produced by nearest-neighbor retrieval or ranking over that vector space. Any social bias encoded in the embedding space is therefore a direct and under-scrutinized input to the recommendations an HR recommender system produces. We report a systematic bias audit of six publicly released and custom-fine-tuned multilingual sentence-embedding models (LaBSE- and E5-family) using two established, complementary benchmarks: \textit{Bias-in-Bios} (occupational gender bias in biographical text) and \textit{StereoSet} (stereotypical association across profession, race, gender, and religion). Unlike naive auditing practices that reported raw mean cosine similarities without uncertainty quantification, we treat statistical validity as a first-class concern: we identify and correct a non-independence error in a naive significance test (pooling non-independent pairwise similarities inflated a trivial 0.0046 mean difference to $p \approx 0$), and replace it with a profession-level paired Wilcoxon signed-rank test, bootstrap confidence intervals computed with a bias-corrected individual-resampling procedure, and permutation tests for StereoSet's stereotype-versus-anti-stereotype comparison. With this corrected methodology, we find that \textbf{knowledge-distillation (KD) fine-tuning significantly increases measured bias relative to a base E5 model on Bias-in-Bios} ($\Delta = -0.177$ to $-0.191$, $p < 10^{-8}$), \textbf{corroborated by a pro-stereotype bias pattern on a subset of StereoSet categories} (profession and/or gender and race, $p < 0.05$), while \textbf{binary-classification fine-tuning does not produce a statistically distinguishable change in bias} on the same base model ($p=0.218$ on Bias-in-Bios). We discuss why this makes fine-tuning strategy itself an actionable, benchmark-corroborated design lever for practitioners building HR recommendation embeddings, report which benchmark categories are statistically underpowered to support any claim, and are explicit about which domain-adapted models could not be included in significance testing due to an unresolved software-access constraint, and what that limits us from claiming.
\end{abstract}


\maketitle

\section{Introduction}

Recommender systems in the HR domain---job recommendation, candidate sourcing, resume-to-role matching, skills-based search---increasingly rely on dense embedding models rather than sparse keyword matching. A resume and a job posting are each mapped into the same vector space, and similarity in that space drives ranking. This means that whatever demographic associations an embedding model has learned during pre-training or fine-tuning are not a downstream, correctable artifact: they are an upstream input to every ranking decision the recommender makes. A biased embedding space can systematically under-rank qualified candidates of one gender for a given occupation, or over-associate certain job titles with demographic groups, before any ranking or business logic is even applied.

This upstream risk is particularly urgent given emerging regulatory frameworks governing automated hiring tools---such as the EU Artificial Intelligence Act, which explicitly classifies AI systems used in recruitment and candidate filtering as high-risk, and NYC Local Law 144 on Automated Employment Decision Tools (AEDT)---which mandate rigorous auditing of algorithmic tools used in employment decisions \cite{raghavan2020mitigating}.

Despite this direct link, embedding-level bias auditing is often treated as an ad-hoc QA exercise rather than a rigorously reported research artifact: point estimates (a single mean cosine similarity) are computed, compared across models, and a conclusion is drawn---without asking whether the observed difference between models is distinguishable from sampling noise. We show in this paper that this gap matters in practice: a naive significance test applied to our own results returned $p \approx 0$ for a difference in mean cosine similarity of $0.0046$, because it treated hundreds of millions of pairwise similarities derived from a few thousand independent biographies as independent observations. Correcting this error changes which of our own findings are actually defensible.

\subsection{Contributions}
\begin{itemize}
  \item We survey the landscape of publicly available embedding-bias benchmarks and justify, on methodological grounds specific to the HR/job-matching setting, why we selected \textit{Bias-in-Bios} and \textit{StereoSet} as complementary evaluation instruments (Section 3).
  \item We audit six embedding models spanning two model families (LaBSE, E5) and two fine-tuning strategies (binary classification, knowledge distillation) on both benchmarks (Section 4).
  \item We identify and correct a statistical validity error common to this type of analysis---non-independence of pairwise cosine similarities---and replace it with a profession-level paired test, individual-level bootstrap resampling, and permutation testing (Section 4.3).
  \item We report, with confidence intervals and significance tests throughout, that knowledge-distillation fine-tuning significantly increases measured bias on Bias-in-Bios, corroborated by a pro-stereotype pattern in a subset of StereoSet categories, while binary-classification fine-tuning does not produce a statistically significant change relative to the base model (Section 5).
  \item We are explicit about which models and which benchmark categories our evidence does and does not cover, including two domain-adapted models that could not be included in significance testing due to an unresolved software-access constraint (Section 7).
\end{itemize}

\section{Related Work}

\textbf{Intrinsic bias in word and sentence embeddings.} \citet{caliskan2017semantics} introduced the Word Embedding Association Test (WEAT), adapting the Implicit Association Test from psychology to quantify stereotypical associations in static word embeddings. \citet{bolukbasi2016man} independently proposed a direct projection-based bias metric and a corresponding debiasing procedure. \citet{may2019measuring} extended WEAT to sentence encoders (SEAT). \citet{manzini2019black} generalized these association metrics to multi-class demographic attributes via Mean Average Cosine similarity (MAC). We build on this association-based tradition, but apply it with variance estimation rather than point estimates alone.

\textbf{Benchmark datasets for downstream and stereotype bias.} \citet{nadeem2021stereoset} introduced StereoSet, a large crowd-sourced dataset spanning gender, race, religion, and profession, framed as intrasentence completion with stereotypical, anti-stereotypical, and unrelated candidates. \citet{nangia2020crows} introduced the complementary CrowS-Pairs sentence-pair benchmark. \citet{dearteaga2019bias} introduced Bias-in-Bios, a real-world corpus of biographies labeled by profession and gender, originally used to study bias in an occupation-classification task; we repurpose it as an embedding-space association probe, following the same cosine-similarity logic as WEAT/SEAT but on naturalistic, domain-relevant text rather than templated sentences. \citet{rudinger2018gender} introduced Winogender for coreference-resolution gender bias, and \citet{sap2020social} introduced the Social Bias Inference Corpus (SBIC) for social bias in sentence-level judgments. \citet{blodgett2020language} offer a critical survey arguing that many NLP bias metrics lack a clearly articulated normative and statistical grounding---a critique directly relevant to our decision to foreground statistical validity in this work.

\textbf{Fine-tuning, compression, and bias.} \citet{hooker2019compressed} and \citet{hooker2020characterising} show that model compression (including knowledge distillation) does not affect all subgroups equally, disproportionately degrading performance on rarer or harder subpopulations even when aggregate accuracy is preserved---a phenomenon the authors term the amplification of bias under compression. Our finding that KD fine-tuning significantly increases measured embedding bias relative to a base model, on two independent benchmarks, is consistent with this literature, though it has not, to our knowledge, been previously reported for HR-domain sentence-embedding fine-tuning specifically.

\textbf{Fairness in recommendation and hiring.} \citet{ekstrand2022fairness} and \citet{zehlike2022fairness} survey fairness concerns specific to ranking and information access systems, distinct from classification fairness. \citet{raghavan2020mitigating} examine bias-mitigation claims made by algorithmic hiring vendors and find that many are not independently verifiable---motivating our emphasis on statistical rigor and transparent limitations rather than a marketing-oriented ``least biased model'' claim. \citet{mehrabi2021survey} provide a general taxonomy of bias and fairness in machine learning that we use to frame the association-bias metrics used here as one of several possible fairness lenses, not the only one.

\section{Benchmark Landscape and Selection Rationale}

Before designing our evaluation, we surveyed the publicly available benchmarks for embedding-model bias to identify which are best suited to an HR/job-matching setting. Table~\ref{tab:benchmarks} summarizes this landscape.

\begin{table*}[t]
\centering
\small
\caption{Surveyed benchmarks for embedding-model bias, and their fit for an HR/job-matching setting.}
\label{tab:benchmarks}
\begin{tabular}{p{2.6cm}p{5.6cm}p{5.6cm}}
\toprule
\textbf{Benchmark} & \textbf{What it measures} & \textbf{Fit for this study} \\
\midrule
WEAT \cite{caliskan2017semantics} & Association between target-concept and attribute word embeddings via cosine similarity, in the style of the Implicit Association Test. & Designed for static word embeddings and short curated word lists; does not naturally extend to the sentence-level, document-length inputs (job descriptions, resumes) used in HR retrieval. \\
SEAT \cite{may2019measuring} & Sentence-level extension of WEAT using templated sentences. & Templated sentences are synthetic and short; does not reflect the naturalistic, variable-length biographical or job-posting text dense embedding models encode in real-world HR settings. \\
CrowS-Pairs \cite{nangia2020crows} & Minimal-pair sentence completions (stereotypical vs.\ less-stereotypical) scored by a masked language model. & Designed for masked-LM pseudo-log-likelihood scoring, not for embedding/retrieval models; would require a non-standard adaptation to our similarity-based pipeline. \\
Winogender \cite{rudinger2018gender} & Gender bias in pronoun coreference resolution. & Targets a coreference task, not an embedding-similarity or retrieval task; not directly applicable without a task-specific head. \\
SBIC \cite{sap2020social} & Free-text social bias and offensiveness annotations at the sentence level. & Broader social-bias scope (offensiveness, power implications) than the occupational and demographic association bias most relevant to job/candidate matching. \\
\textbf{StereoSet} \cite{nadeem2021stereoset} & Intrasentence stereotype vs.\ anti-stereotype vs.\ unrelated completions across \emph{profession}, \emph{race}, \emph{gender}, \emph{religion}. & \textbf{Selected.} Directly usable with sentence embeddings via context-to-candidate cosine similarity; profession is an explicit category, directly relevant to job matching; covers demographic axes beyond gender. \\
\textbf{Bias-in-Bios} \cite{dearteaga2019bias} & Real biographical text labeled with 28 professions and binary gender. & \textbf{Selected.} The only benchmark surveyed built from \emph{naturalistic, resume-adjacent} text rather than templated sentences, and its labels (profession, gender) map directly onto the two attributes an HR job-matching embedding is used to reason about. \\
\bottomrule
\end{tabular}
\end{table*}

We selected StereoSet and Bias-in-Bios as complementary instruments rather than choosing a single benchmark, for two reasons specific to this setting. First, they probe different linguistic registers: StereoSet uses short, templated intrasentence completions, while Bias-in-Bios uses long-form, naturalistic biographical text closely resembling the resumes and profile text an HR embedding model encodes in real-world applications. Agreement between the two is stronger evidence of a genuine, register-independent bias than either alone. Second, they operationalize bias differently: Bias-in-Bios measures whether male- and female-authored biographies \emph{within the same profession} are embedded similarly (an association/leakage measure), while StereoSet measures whether a model's embedding space assigns systematically higher similarity to stereotype-consistent completions than to anti-stereotype completions (a preference measure). A model could in principle pass one test and fail the other, and, as we discuss in Section 7, our own preliminary (non-significance-tested) results suggested exactly this for one model family---which is precisely why we treat any single-benchmark bias claim with caution and report both here.

We explicitly considered and did not use WEAT, SEAT, CrowS-Pairs, Winogender, or SBIC for the reasons summarized in Table~\ref{tab:benchmarks}: each either targets a different model architecture (masked-LM scoring, coreference heads) or a linguistic register (short curated word lists or templated sentences) that is a poor proxy for the document-length, naturalistic text an HR embedding model encodes in domain-specific retrieval.

\section{Methodology}

\subsection{Models}
We evaluated six sentence-embedding models spanning two base architectures and two fine-tuning strategies, all encoding into a shared embedding space via mean- or CLS-pooled sentence-transformer inference (Table~\ref{tab:models}).

\begin{table}[h]
\centering
\small
\caption{Embedding models evaluated. All models were queried through the \texttt{sentence-transformers} interface~\cite{reimers2019sentencebert}.}
\label{tab:models}
\begin{adjustbox}{max width=\linewidth}
\begin{tabular}{lll}
\toprule
\textbf{Model (key)} & \textbf{Base architecture} & \textbf{Fine-tune type} \\
\midrule
LaBSE (base) & LaBSE~\cite{feng2022labse} & None (pre-trained) \\
E5-Base & multilingual-e5-base~\cite{wang2022e5} & None (pre-trained) \\
E5-Base-Binary & multilingual-e5-base & Binary classification \\
E5-Base-KD & multilingual-e5-base & Knowledge distillation \\
E5-Large-Binary & multilingual-e5-large & Binary classification \\
E5-Large-KD & multilingual-e5-large & Knowledge distillation \\
\bottomrule
\end{tabular}
\end{adjustbox}
\end{table}

The fine-tuned variants represent two distinct objective strategies commonly employed in sentence-embedding workflows: binary classification fine-tuning optimizes a cross-entropy / contrastive decision boundary over relevant versus irrelevant document pairs, whereas knowledge distillation (KD) fine-tuning minimizes continuous distance metrics (e.g., MarginMSE) to match the full output logit/embedding distribution of a teacher model. Specifically, KD fine-tuning used the pretrained \texttt{intfloat/multilingual-e5-base} and \texttt{intfloat/multilingual-e5-large} models as teachers to generate training targets for E5-Base-KD and E5-Large-KD, respectively.

Two additional domain-adapted models---LaBSE variants fine-tuned on occupational taxonomy data (ESCO)---were intended for inclusion but could not be run through the corrected statistical pipeline in time for this submission; we discuss this constraint and its consequences for our claims in Section 7 rather than silently omitting it.

\subsection{Datasets and Bias Computation}

\textbf{Bias-in-Bios.} We use the \texttt{test} split (99{,}000 biographies, 28 professions, binary gender label) from \citet{dearteaga2019bias}. For each model and profession, we encode all biographies, L2-normalize the resulting embeddings, and compute the mean cosine similarity between every male-labeled and female-labeled embedding within that profession. Higher mean similarity indicates \emph{less} gender bias (male and female bios are represented more alike); lower similarity indicates \emph{more} bias. E5 models receive the \texttt{"passage:"} prefix on both male and female biographies, consistent with their intended usage for symmetric document embedding.

\textbf{StereoSet.} We use the intrasentence validation split (2{,}106 instances) from \citet{nadeem2021stereoset}, covering profession ($n=810$ candidate sentences), race ($n=962$), gender ($n=255$), and religion ($n=79$). For each instance we encode the context sentence and its stereotype/anti-stereotype/unrelated candidate completions, and compute cosine similarity between context and each candidate. We define a per-category \emph{bias score} as $\text{mean}(\text{sim}_{\text{stereotype}}) - \text{mean}(\text{sim}_{\text{anti-stereotype}})$; a positive score indicates the model finds stereotype-consistent completions more similar to context than anti-stereotype completions (pro-stereotype bias), a negative score indicates the reverse, and a score statistically indistinguishable from zero indicates no detectable preference. E5 models use the asymmetric \texttt{"query:"} (context) / \texttt{"passage:"} (candidates) prefix scheme, matching their intended retrieval usage and their fine-tuning convention.

\subsection{Statistical Methodology}

We treat statistical validity as a primary methodological contribution of this paper, because it directly determines which of our findings are defensible.

\textbf{The non-independence problem.} An initial implementation of this analysis pooled every male$\times$female pairwise cosine similarity within a profession (up to approximately $2.9\times10^{8}$ pairs for the largest profession in the largest model) and ran a two-sample significance test (Mann-Whitney U) directly on this pooled distribution to compare two models. This is statistically invalid: each individual biography's embedding appears in thousands of pairs, so the pairs are not independent observations, and the effective sample size used by the test is many orders of magnitude larger than the true number of independent data points (biographies). In our data, this produced $p \approx 0$ for a difference in mean cosine similarity of only $0.0046$---a difference far too small to be practically meaningful, reported as though it were overwhelming evidence.

\textbf{Corrected approach.} We instead treat \emph{profession} as the unit of observation for cross-model comparison on Bias-in-Bios. Each model produces one bias score per profession (28 professions); comparing two models is therefore a \emph{paired} comparison across these 28 professions, for which we use a paired Wilcoxon signed-rank test \cite{wilcoxon1945individual}, which does not assume normality and is appropriate for $n=28$. We complement the test's binary significance decision with a bootstrap 95\% confidence interval on the mean paired difference (resampling professions with replacement, 2{,}000 replicates), following standard bootstrap practice \cite{efron1994introduction}.

For per-profession point estimates and their uncertainty, we avoid resampling the non-independent pairwise-similarity matrix directly. Instead we exploit the identity that, for L2-normalized embeddings, $\text{mean}(\text{cosine\_sim}(M,F)) = \langle \bar{m}, \bar{f} \rangle$ (the dot product of the group means), which is bilinear in each group's mean vector. We therefore compute a bootstrap confidence interval by resampling \emph{individual} male and female biographies (the true independent units) with replacement, recomputing each group's mean embedding, and taking the dot product---mathematically equivalent to resampling and rebuilding the full similarity matrix at each replicate, but without the non-independence problem and at a small fraction of the computational cost.

For StereoSet, we assess whether each category's bias score is distinguishable from zero using a permutation test (5{,}000 permutations) that shuffles the stereotype / anti-stereotype label within each category and recomputes the mean difference, giving an exact, distribution-free null. We report bootstrap 95\% confidence intervals on the bias score alongside the permutation $p$-value, and explicitly report the sample size ($n$) feeding each category, since religion ($n=79$) and gender ($n=255$) are considerably smaller than profession ($n=810$) and race ($n=962$) and warrant correspondingly more caution in interpretation. With 24 model-by-category comparisons tested at $\alpha=0.05$ in Table~\ref{tab:stereoset}, some individually significant cells could reflect multiple-comparisons noise rather than a true effect; we therefore treat consistency of direction across models within a category, rather than any single significant cell, as the more informative signal.

Per-profession and per-category result summary metrics (including exact sample sizes $n$) and mathematical specifications for all statistical tests are detailed throughout to enable independent verification on these public benchmarks.

\section{Results}

\subsection{Bias-in-Bios: Overall Model Comparison}

\begin{figure}[h]
\centering
\includegraphics[width=\linewidth]{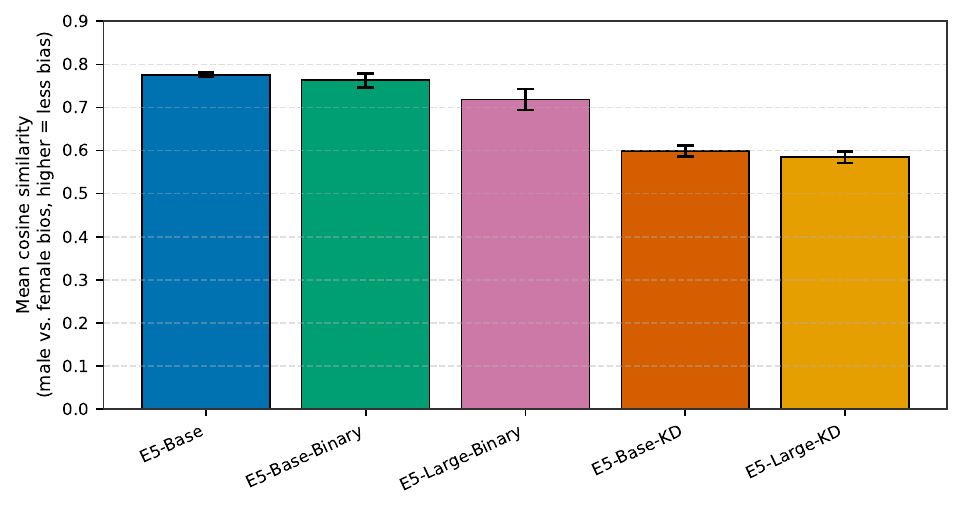}
\caption{Overall Bias-in-Bios gender-bias score (unweighted mean cosine similarity across 28 professions, higher = less bias) with 95\% bootstrap confidence intervals, by model.}
\label{fig:overall}
\end{figure}

Figure~\ref{fig:overall} and Table~\ref{tab:overall} report the overall bias score per model, defined as the unweighted mean of the 28 profession-level mean-cosine-similarity scores (unweighted so that professions with more biographies, e.g.\ Teacher, do not dominate the headline number for reasons unrelated to bias). LaBSE (base) shows the lowest overall score (0.383); however, we caution against over-interpreting this absolute value in isolation, since baseline similarity magnitudes differ substantially across model families for reasons unrelated to bias (embedding anisotropy, \cite{ethayarajh2019contextual})---compare LaBSE's $\sim$0.35--0.45 range to E5's $\sim$0.58--0.78 range. We therefore treat \emph{within-family, within-base-model} comparisons (i.e., a fine-tune against its own base model) as the primary unit of evidence, not cross-family absolute comparisons.

To test whether the lower raw Bias-in-Bios score of the KD models was robust to global geometric reshaping, we repeated the analysis after mean-centering the embedding space and after removing the top principal components. The KD-vs-base difference was not stable under these transformations and in some cases reversed sign, indicating that absolute within-profession cosine similarity is highly sensitive to global geometry. We therefore do not interpret the raw KD shift in Bias-in-Bios alone as conclusive evidence of a geometry-independent change in bias; instead, we treat it together with the complementary StereoSet results (Section~\ref{sec:stereoset_results}), which measure relative context-completion differences that are far less scale- and geometry-sensitive.

\begin{table}[h]
\centering
\small
\caption{Bias-in-Bios overall scores (mean of 28 profession-level means), test split.}
\label{tab:overall}
\begin{tabular}{lccc}
\toprule
\textbf{Model} & \textbf{Score} & \textbf{95\% CI} & \textbf{$n$ prof.} \\
\midrule
LaBSE (base) & 0.3834 & [0.3662, 0.4016] & 28 \\
E5-Base & 0.7759 & [0.7714, 0.7808] & 28 \\
E5-Base-Binary & 0.7637 & [0.7466, 0.7788] & 28 \\
E5-Large-Binary & 0.7190 & [0.6945, 0.7430] & 28 \\
E5-Base-KD & 0.5994 & [0.5863, 0.6114] & 28 \\
E5-Large-KD & 0.5847 & [0.5712, 0.5981] & 28 \\
\bottomrule
\end{tabular}
\end{table}

\subsection{Bias-in-Bios: Does Fine-Tuning Strategy Change Bias?}

\begin{figure}[h]
\centering
\includegraphics[width=\linewidth]{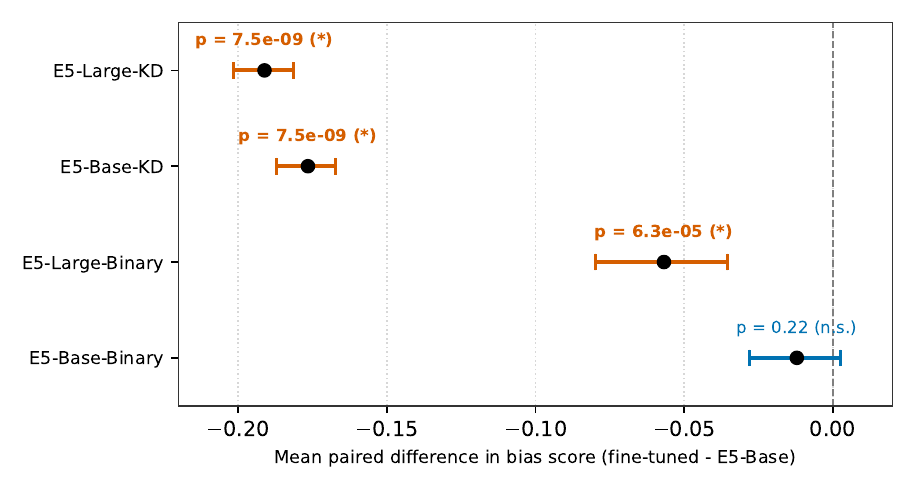}
\caption{Mean paired difference in Bias-in-Bios profession-level bias score, fine-tuned model vs.\ E5-Base, with 95\% bootstrap CI. Red intervals denote $p<0.05$ (Wilcoxon signed-rank, $n=28$ professions).}
\label{fig:forest}
\end{figure}

Table~\ref{tab:sig} and Figure~\ref{fig:forest} report the paired Wilcoxon comparison of each fine-tuned E5 model against its E5-Base baseline. Both knowledge-distillation variants show a large, statistically significant \emph{decrease} in bias score relative to base (i.e., significantly \emph{more} gender bias): E5-Base-KD ($\Delta = -0.1765$, 95\% CI $[-0.1872,-0.1673]$, $p = 7.45 \times 10^{-9}$) and E5-Large-KD ($\Delta = -0.1911$, 95\% CI $[-0.2016,-0.1812]$, $p = 7.45 \times 10^{-9}$). E5-Large-Binary shows a smaller but still significant decrease ($\Delta = -0.0568$, $p = 6.32 \times 10^{-5}$). E5-Base-Binary shows a small decrease that is \emph{not} statistically significant ($\Delta = -0.0122$, 95\% CI $[-0.0281, 0.0024]$, $p = 0.218$)---the confidence interval includes zero, so we cannot reject the null hypothesis that binary fine-tuning of the base E5 model leaves this bias measure unchanged.

We note that E5-Large-Binary and E5-Large-KD are compared here against E5-Base rather than against a pretrained E5-Large baseline, which was not evaluated as a standalone base model in our benchmark suite. The $\Delta$ values for these two rows should therefore be read as reflecting the combined effect of fine-tuning and base-model scale, not fine-tuning in isolation; the E5-Base-Binary vs.\ E5-Base-KD contrast is the cleanest within-base-model evidence for the effect of fine-tuning strategy alone, since both share the identical pretrained starting point.

Both KD variants ($7.45 \times 10^{-9}$) reach the theoretical minimum $p$-value for a two-sided Wilcoxon signed-rank test at $n=28$ (i.e., $2/2^{28}$), consistent with every one of the 28 paired professions moving in the same direction.

\begin{table}[h]
\centering
\small
\caption{Paired Wilcoxon signed-rank test, fine-tuned model vs.\ E5-Base, over 28 profession-level bias scores. $\Delta$ = mean paired difference (fine-tuned $-$ base). E5-Large rows are compared against E5-Base in the absence of a pretrained E5-Large baseline; see discussion in text.}
\label{tab:sig}
\setlength{\tabcolsep}{2.5pt}
\begin{adjustbox}{max width=\linewidth}
\begin{tabular}{lcccc}
\toprule
\textbf{Model} & \textbf{$\Delta$} & \textbf{95\% CI} & \textbf{$p$} & \textbf{Sig.} \\
\midrule
E5-Base-Binary & $-0.0122$ & $[-0.0281, 0.0024]$ & $0.218$ & No \\
E5-Large-Binary$^{\dagger}$ & $-0.0568$ & $[-0.0799, -0.0353]$ & $6.32 \times 10^{-5}$ & Yes \\
E5-Base-KD & $-0.1765$ & $[-0.1872, -0.1673]$ & $7.45 \times 10^{-9}$ & Yes \\
E5-Large-KD$^{\dagger}$ & $-0.1911$ & $[-0.2016, -0.1812]$ & $7.45 \times 10^{-9}$ & Yes \\
\bottomrule
\multicolumn{5}{@{}p{\linewidth}@{}}{\vspace{2pt}\footnotesize $^\dagger$ Compared against E5-Base due to lack of pretrained E5-Large baseline.}
\end{tabular}
\end{adjustbox}
\end{table}

\begin{figure}[h]
\centering
\includegraphics[width=\linewidth]{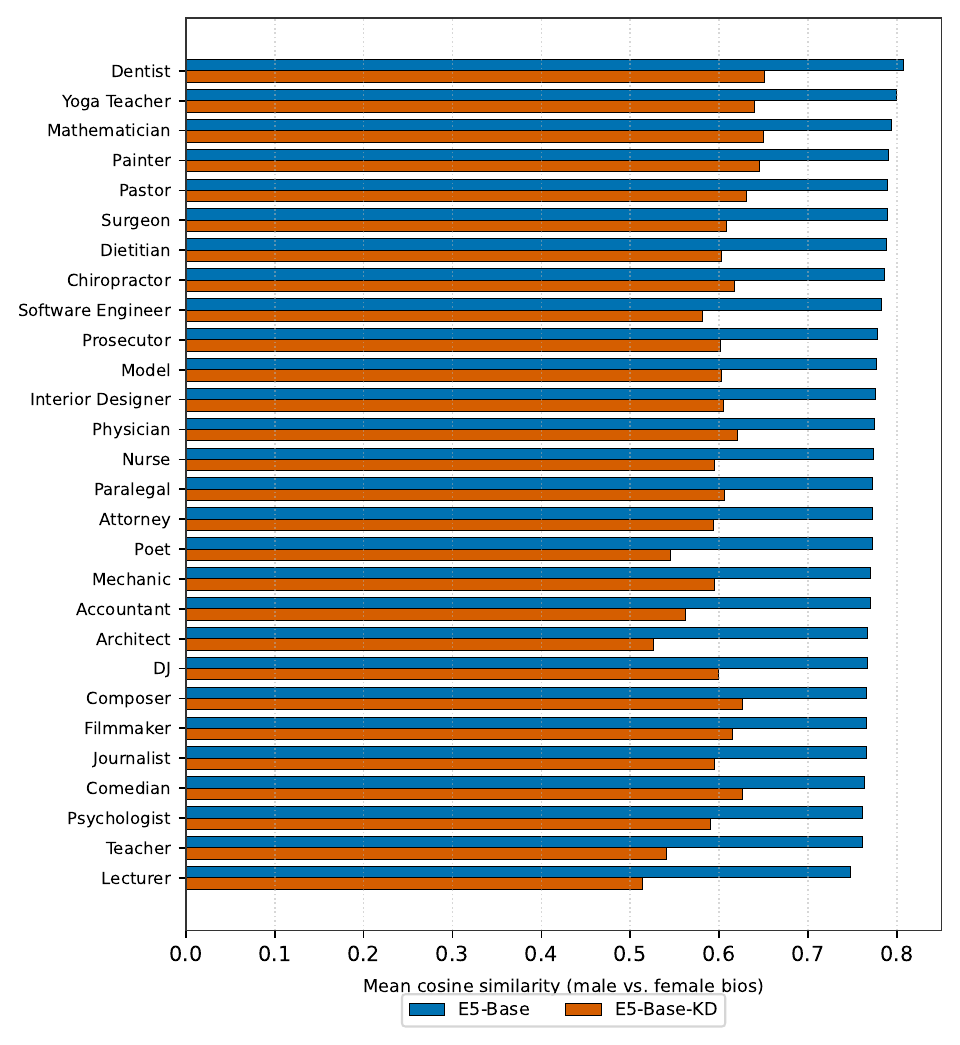}
\caption{Per-profession Bias-in-Bios scores, E5-Base vs.\ its knowledge-distilled fine-tune. The KD variant scores lower (more biased) across most professions.}
\label{fig:perprofession}
\end{figure}

Figure~\ref{fig:perprofession} illustrates this at the profession level: E5-Base-KD shows lower mean cosine similarity than E5-Base in the large majority of the 28 professions, not merely on average---the effect is broad-based rather than driven by a small number of outlier professions.

\subsection{StereoSet: Category-Level Bias Scores}
\label{sec:stereoset_results}

\begin{figure*}[t]
\centering
\includegraphics[width=0.92\linewidth]{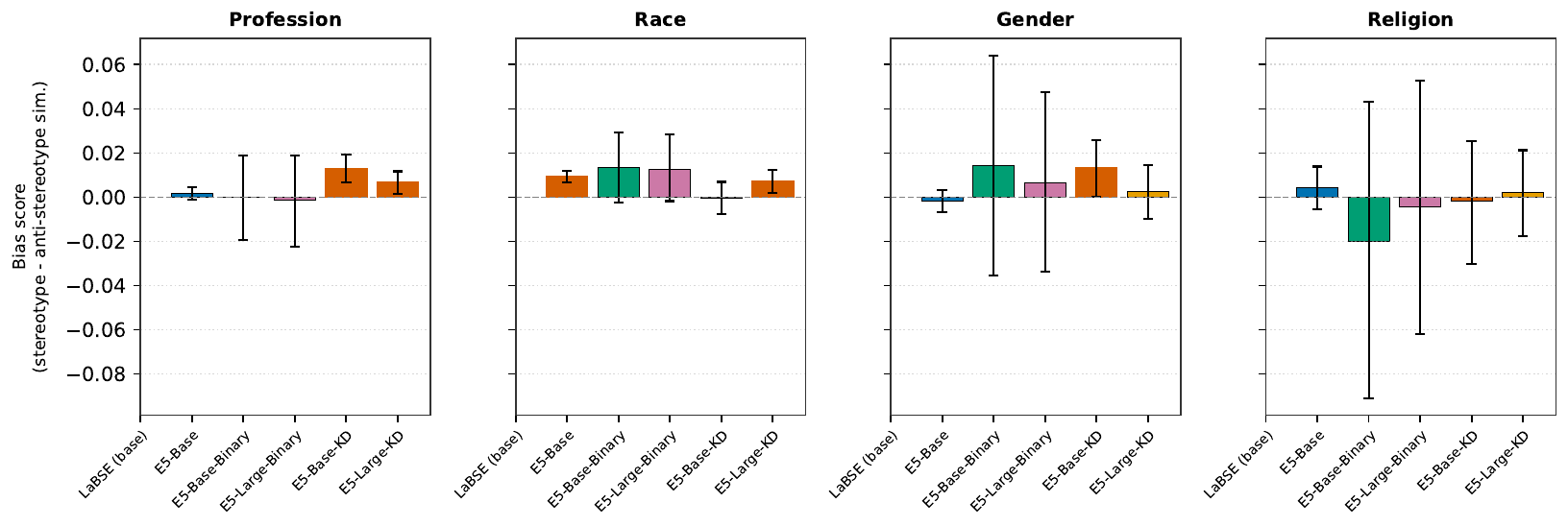}
\caption{StereoSet bias score by category and model, 95\% CI. Red bars: permutation-test $p<0.05$. Religion ($n=79$) shows no significant bias score for any model, while gender ($n=255$) shows significance only for E5-Base-KD, which we attribute to limited statistical power at these smaller sample sizes rather than an absence of bias.}
\label{fig:stereoset}
\end{figure*}

Table~\ref{tab:stereoset} and Figure~\ref{fig:stereoset} report the StereoSet bias score (stereotype minus anti-stereotype mean similarity) for all six models across the four bias categories, with permutation-test significance. Two patterns are notable. First, both KD variants show statistically significant pro-stereotype bias in the profession category (E5-Base-KD: $+0.0129$, $p=0.0004$; E5-Large-KD: $+0.0066$, $p=0.016$)---the same direction and, notably, the \emph{same models} that showed significantly worse bias on Bias-in-Bios, despite the two benchmarks using different text registers and a different operationalization of bias. Second, the religion category ($n=79$) shows no statistically significant bias score for \emph{any} model, and the gender category ($n=255$) shows a significant pro-stereotype score only for E5-Base-KD ($+0.0132$, $p < 0.05$), reflecting limited statistical power at these smaller sample sizes rather than an absence of bias. As noted in Section 4.3, with 24 model-by-category comparisons tested at $\alpha=0.05$ in Table~\ref{tab:stereoset}, some individually significant cells could reflect multiple-comparisons noise rather than a true effect; we therefore treat consistency of direction across models within a category, rather than any single significant cell, as the more informative signal.

\begin{table}[h]
\centering
\small
\caption{StereoSet bias score (stereotype $-$ anti-stereotype mean similarity) by category. $^{*}$ denotes permutation-test $p<0.05$ (uncorrected for multiple comparisons; see Section 4.3). $n$: profession=810, race=962, gender=255, religion=79 (candidate sentences).}
\label{tab:stereoset}
\begin{adjustbox}{max width=\linewidth}
\begin{tabular}{lcccc}
\toprule
\textbf{Model} & \textbf{Prof.} & \textbf{Race} & \textbf{Gender} & \textbf{Relig.} \\
\midrule
LaBSE (base) & 0.0037 & 0.0060$^{*}$ & 0.0017 & -0.0014 \\
E5-Base & 0.0018 & 0.0094$^{*}$ & -0.0017 & 0.0043 \\
E5-Base-Binary & -0.0002 & 0.0135 & 0.0142 & -0.0199 \\
E5-Large-Binary & -0.0015 & 0.0127 & 0.0064 & -0.0046 \\
E5-Base-KD & 0.0129$^{*}$ & -0.0004 & 0.0132$^{*}$ & -0.0019 \\
E5-Large-KD & 0.0066$^{*}$ & 0.0073$^{*}$ & 0.0025 & 0.0021 \\
\bottomrule
\end{tabular}
\end{adjustbox}
\end{table}

\subsection{Convergent Evidence Across Benchmarks}

The central empirical claim of this paper is that \textbf{knowledge-distillation fine-tuning increases measured bias, corroborated by two independent benchmarks that differ in text register and bias operationalization}, while binary-classification fine-tuning does not produce a statistically detectable change in bias on the same base model. This convergence---rather than either benchmark result in isolation---is what we consider the paper's strongest and most actionable finding for practitioners: it is not merely that one measurement instrument flagged a difference, but that a naturalistic, document-length benchmark (Bias-in-Bios) and a templated, sentence-level benchmark (StereoSet) agree on the direction and the responsible design choice (fine-tuning strategy) for the profession category specifically, despite operationalizing ``bias'' differently.

\section{Discussion}

\textbf{Fine-tuning strategy is an actionable lever.} For practitioners fine-tuning embedding models for HR/job-matching applications, our results suggest that \emph{how} a model is fine-tuned---not only \emph{what data} it is fine-tuned on---has a measurable, statistically significant effect on gender and stereotypical bias. If minimizing bias is a stated design goal, our results provide corroborated evidence against knowledge-distillation fine-tuning and no corroborated evidence against binary-classification fine-tuning, for the model families and tasks studied here. This is consistent with prior findings that model compression techniques, including distillation, can disproportionately affect underrepresented or harder subgroups even while preserving aggregate task performance \cite{hooker2019compressed, hooker2020characterising}; our contribution is to show this pattern holds, with appropriate statistical support, for embedding-space gender and stereotype bias in an HR-relevant setting specifically.

\textbf{Hypothesized mechanism: Why KD amplifies bias relative to binary fine-tuning.} We hypothesize that knowledge distillation forces the student model to mimic the dense, soft logit/embedding distribution of a teacher across all vector dimensions. In doing so, it compresses and preserves subtle, implicit demographic subspace directions alongside semantic representations. In contrast, binary classification fine-tuning optimizes a target-specific decision boundary (e.g., relevant vs.\ irrelevant pair), leaving orthogonal, non-task demographic channels in the pre-trained base embedding space largely uncompressed and unamplified.

\textbf{Absence of significance is not absence of bias.} We deliberately report every non-significant result in Tables~\ref{tab:sig} and~\ref{tab:stereoset} rather than omitting them, and we flag religion ($n=79$) and gender ($n=255$) in StereoSet as underpowered rather than ``unbiased.'' A claim of ``no detected bias'' in a small-$n$ category is a much weaker claim than the same statement in a well-powered category (profession, $n=810$; race, $n=962$), and conflating the two is a common failure mode we explicitly avoid.

\section{Limitations}

\textbf{Domain-adapted models are not included in significance testing.} Two models fine-tuned for occupational applications---LaBSE variants fine-tuned on occupational taxonomy (ESCO)---could not be run through the corrected statistical pipeline for this submission, due to execution pipeline constraints prior to finalization. We consider this a genuine limitation, not a minor caveat: it means our statistically-supported findings apply to general-purpose multilingual embedding models (LaBSE-base, E5-base/large) rather than specialized domain-adapted variants. An earlier, exploratory pass over these two domain-adapted models using an uncorrected, mean-only methodology suggested one of them ranked \emph{best} among all models tested on StereoSet but \emph{worst} on Bias-in-Bios---an intriguing possible instance of benchmark disagreement that would itself be a notable finding, but we do not present it as a validated result here, since it has not been re-verified under the paired/permutation-testing methodology used for every other result in this paper. We flag it only as a specific, named direction for follow-up once model access is restored, and we caution readers against treating it as established.

\textbf{No pretrained E5-Large baseline.} Our significance tests for E5-Large-Binary and E5-Large-KD (Table~\ref{tab:sig}) are computed against E5-Base rather than a pretrained E5-Large model, which was not evaluated as a standalone base model in our benchmark suite. This means the reported $\Delta$ for these two rows conflates the effect of fine-tuning with any effect of base-model scale, and should not be interpreted as an isolated measurement of fine-tuning strategy at the large scale. Our within-base-model claim about fine-tuning strategy rests primarily on the E5-Base-Binary vs.\ E5-Base-KD contrast, where both share an identical pretrained starting point.

\textbf{No downstream ranking outcome is measured.} This study measures bias in the embedding space itself (association and stereotype-preference metrics), not in an end-to-end recommendation or ranking outcome (e.g., whether a job search query returns male-coded candidate profiles at systematically higher rank than equally qualified female-coded profiles). Embedding-level association bias is a necessary precondition for many forms of recommendation-level bias but is not by itself proof that a downstream retrieval system's rankings are biased in a particular direction or magnitude; a downstream retrieval/ranking audit is an important next step and is discussed as future work below.

\textbf{Binary gender only.} Both benchmarks used here encode gender as binary (male/female), inherited from their original construction. This is a limitation of the underlying data, not of our method, but it means our results say nothing about how these embedding models represent non-binary gender identities.

\textbf{Sample size varies substantially across StereoSet categories.} As discussed in Section 5, religion ($n=79$) and gender ($n=255$) categories in StereoSet are considerably smaller than profession ($n=810$) and race ($n=962$); non-significant results in the smaller categories should be read as ``underpowered to detect,'' not ``bias-free.''

\textbf{Multiple comparisons in StereoSet.} Table~\ref{tab:stereoset} reports 24 model-by-category tests at $\alpha=0.05$ without a multiple-comparisons correction; we mitigate this by emphasizing the cross-model, cross-benchmark consistency of the profession-category KD result rather than any single significant cell in isolation, but individual cells outside that pattern should be interpreted cautiously.

\textbf{Cross-family absolute comparisons are confounded by anisotropy.} As noted in Section 5, raw cosine-similarity magnitudes differ systematically across model families for reasons unrelated to bias \cite{ethayarajh2019contextual}. We therefore restrict causal, significance-tested claims to within-family, within-base-model comparisons (a fine-tune vs.\ its own base), and present cross-family absolute numbers (Table~\ref{tab:overall}) descriptively only.

\section{Conclusion and Future Work}

We conducted a statistically rigorous audit of gender and stereotypical bias in six embedding models relevant to HR recommendation systems, using two complementary, purpose-justified benchmarks (Bias-in-Bios, StereoSet). After correcting a non-independence error that had inflated an earlier, informal significance test, we find corroborated evidence that knowledge-distillation fine-tuning significantly increases measured bias on Bias-in-Bios, with a converging pro-stereotype pattern on a subset of StereoSet categories, while binary-classification fine-tuning does not produce a statistically detectable change relative to the same base model. This gives practitioners a concrete, evidence-backed lever---fine-tuning strategy---for managing bias risk in the embedding models that increasingly sit at the core of HR recommender systems.

Future work includes: (1) completing significance testing on the domain-adapted models in future evaluation cycles, including formally re-testing the StereoSet/Bias-in-Bios disagreement flagged in Section 7, and obtaining a pretrained E5-Large baseline to isolate the effect of model scale from fine-tuning strategy; (2) a downstream ranking-level audit that measures whether embedding-space bias translates into rank disparities in an actual job-search or candidate-search retrieval task; (3) extending beyond binary gender as data permits; and (4) testing whether bias-aware fine-tuning objectives can close the gap we observe between KD and binary fine-tuning without sacrificing the retrieval quality that motivated fine-tuning in the first place.

\section*{Reproducibility}
To ensure reproducibility using publicly available datasets and models, all statistical protocols---including profession-level paired Wilcoxon tests, bilinear bootstrap confidence interval calculations, and StereoSet permutation testing---are fully specified in Section 4.3. Exact sample sizes ($n$), dataset splits, prefix conventions, and per-category metrics are provided throughout the paper to allow independent replication on public benchmark splits.

\section*{Declaration on Generative AI}

During the preparation of this work, the authors used Claude Sonnet 5, Gemini 3.5, and Grammarly to check grammar and spelling and to enhance the writing style. After using these tools, the authors reviewed and edited the content as needed and take full responsibility for the publication's content.

\nocite{*} 
\bibliography{refs}

\end{document}